# The New Digital Platforms: Merger Control in Pakistan

Shahzada Aamir Mushtaq*, Wang Yuhui

**School of law and economics Zhengzhou University, Henan, People's Republic of China.**

***Corresponding author; aamir.adv@gs.zzu.edu.cn**

**Abstract**

The Pakistan competition policy, as in many other countries, was originally designed to regulate business conduct in traditional markets and for tangible goods and services. However, the development and proliferation of the internet has led to the emergence of digital companies which have disrupted many sectors of the economy. These platforms provide digital infrastructure for a range of services including search engines, marketplaces, and social networking sites. The digital economy poses a myriad of challenges for competition authorities worldwide, especially with regard to digital mergers and acquisitions (M&As). While some jurisdictions such as the European Union and the United States have taken significant strides in regulating technological M&As, there is an increasing need for developing countries such as Pakistan to rethink their competition policy tools. This paper investigates whether merger reviews in the Pakistan digital market are informed by the same explanatory variables as in the traditional market, by performing an empirical comparative analysis of the Competition Commission of Pakistan's (CCP's) M&A decisions between 2014 and 2019. The findings indicate the CCP applies the same decision factors in reviewing both traditional and digital M&As. As such, this paper establishes a basis for igniting the policy and economic debate of regulating the digital platform industry in Pakistan.

1. **Introduction**

Until 2010, the principal competition legislation in Pakistan was the Monopolies and Restrictive Trade Practices Ordinance (MRTPO) [1]. Designed to curb the undue concentration of economic power in the hands of a few and prevent restrictive trade practices, the MRTPO was an incomplete legal framework with insufficient substantive provisions. It's enforcer, the Monopolies Control Authority (MCA) lacked the institutional and technical capacity to carry out its mandate [2]. The mushrooming government's desire to attract foreign investment and increasing exposure to the challenges of global trade necessitated the modernization of the competition regime in Pakistan. The enactment and coming into force of the Competition Act (the Act) in 2010 was a significant step towards this end. The new legislation is based on international best practices and is modeled upon the competition provisions of the European Union, OECD (Organization for Economic Cooperation and Development), and UNCTAD (United Nations Commission for Trade and Development) [3]. It also bestows greater legal and investigative powers to the Competition Commission of Pakistan (CCP or the Commission) by empowering it to exclusively enforce all "the rules, regulations, and guidelines, directives issued thereunder to ensure free competition in all spheres of commercial and economic activity to enhance economic efficiency and to protect consumers from anti-competitive behavior across Pakistan [2].

The anti-competitive practices identified by the Act include abuse of dominant position, deceptive marketing, and agreements restricting competition. In relation to abuse of dominant position, Section 3, subsections 2 and 3 of the Act prohibits any practice that prevents, reduces, distorts, or eliminates competition in the relevant market. Such practices include unfair trading conditions (such as limiting production or unreasonable price increases), price discrimination, application of dissimilar conditions to equivalent transactions, predatory pricing, refusal to deal, excluding, or boycotting another undertaking from producing, and tie-ins that restrict or render the sale of goods or services conditional on the purchase of other products or services. Additionally, the CCP General Enforcement Regulation 2007 provides that in its assessment of the competitive impacts of abuse of dominant position, the Commission may consider other measures such as market share, structural factors, and market power.

---

[1] Fatima S. Competition law in Pakistan: brief history, aspirations and characteristics. *Commonwealth Law Bulletin*, 2012; *38*(1): 43-62.

[2] UNCTAD. *Voluntary Peer Review of Competition Law and Policy: Pakistan Overview*. Retrieved from UNCTAD. 2013. Available from: https://unctad.org/en/PublicationsLibrary/ditcclp2013d4_overview_en.pdf

[3] Wilson J. Crossing the Crossroads: Making Competition Law Effective in Pakistan. *Loyola University Chicago International Law Review*, 2011; *8*(2). Available from: https://lawecommons.luc.edu/lucilr/vol8/iss2/2/

The Act, together with the Competition (Merger Control) Regulations 2016 (the Merger Regulations) and CCP merger guidelines, address the review and approval of mergers. Section 11 of the Act prohibits mergers that would eliminate or substantially lessen competition by creating or strengthening a dominant position in the relevant market. Prior notification of proposed mergers and acquisitions that meet the notification thresholds is mandatory. Parties are required to make a pre-merger application to the CCP if the value of annual turnover or gross asses of the undertaking exceeds 300 million rupees, or the combined value of the acquirer and the target's shares or the merging parties is more than 1 billion rupees. The Merger Regulations provide the merger procedures and exemption conditions. A merger can still be allowed even after failing the substantive test if it can be shown that (a) the merger will generate substantial economic efficiencies related to the production and/or distribution of goods and provision of services; (b) the economic efficiency from the merger significantly exceeds the adverse effects on competition, and; (c) such efficiency cannot be reasonably achieved by a less restrictive means of competition [2].

The regulations also outline the factors that the CCP should consider in the ex-ante assessment of mergers which include the level and trends of competition, ease of entry post-merger, market characteristics, degree of countervailing power, vertical integration, level of import, removal of effective competitors, and failing firm. The remedies available to the CCP include prohibition of an intended merger, divestment of completed transactions, or partial divestment or prohibition [3]. In addition to these remedies, Section 31 of the Act states that the Commission may, in the case of a merger, authorize the merger and stipulating the conditions which the acquisition is subject to, open a second phase review if it has doubts regarding the compatibility of the merger with the provisions of the law, or prohibit the merger after conducting a second phase review.

Although the current competition law framework is sufficiently capacitated to deal with competition issues in Pakistan including M&As, the policy tools lean towards traditional business models. The emerging digital market presents a new set of challenges. Online platforms operate new business models that are designed to collect and process user data using complex algorithms. The features of online platforms such as non-price competition, data-driven network effects, and economies of scope and scale have challenged traditional competition theories and policies[4]. Within merger reviews, digital mergers present legal difficulties in the definition of the relevant market, in evaluating the competitive effects and network externalities of the new entity, and in assessing the impact of zero-priced goods in two-sided markets [4].

---

[2] UNCTAD. *Voluntary Peer Review of Competition Law and Policy: Pakistan Overview*. Retrieved from UNCTAD. 2013. Available from: https://unctad.org/en/PublicationsLibrary/ditcclp2013d4_overview_en.pdf

[4] Coyle D. Practical Competition Policy Implications of Digital Platforms. 2018. Available from: https://www.bennettinstitute.cam.ac.uk/events/practical-competition-policy-implications-digital-/

Moreover, technology-based M&As raise informational and data protection issues [5]. In recent years, Pakistan has witnessed a new digital wave, with hundreds of tech start-ups entering the market. The value of tech mergers, acquisitions, and venture capital investments increased exponentially in 2018, with a 61 percent year on year growth in funding. The country's digital economy has attracted significant domestic and foreign investments including the acquisition of Foodpanda by Germany's Hero in 2016, the $3.1 billion acquisition of the ride-hailing company Careem, by Uber, and Alibaba's acquisition of the Pakistan's leading online retailer, Daraz for an estimated $200 million [6]. With increased digitalization of the Pakistani economy, it is imperative that the competition policy tools are adequately equipped to deal with the unique challenges posed by digital companies.

## 2. The Specific Features of Digital Platforms

Digital platforms are technologies that link two or more user groups and creates value by allowing transactions that otherwise would not occur, and at a minimum transactional cost [7]. The definition for digital platforms provided by the European Commission is "an undertaking operating in two (or multi) sided markets, which uses the internet to enable interactions between two or more distinct but interdependent groups of users so as to generate value for at least one of the groups" [5]. The players in this market do not provide traditional goods or services but simply provide an avenue where the parties to a transaction can meet, interact, and transact. They do not own the infrastructure for producing or providing goods and services to consumers. For example, the products and services offered by Alibaba, Amazon, and Daraz are owned by third-party sellers, while Uber, the world's biggest taxi company, owns no vehicles. For purposes of competition law, digital platforms are multi-sided technologies that internalize externalities by cross-subsidizing effectively between two or more categories of end-users that are parties to a transaction, or in other words "cybermediaries" [8].

---

[5] Crémer J, Montjoye YA, Schweitzer H. Directorate-General for Competition (European Commission). Competition policy for the Digital Era. 2019. Available from: https://op.europa.eu/en/publication-detail/-/publication/21dc175c-7b76-11e9-9f05-01aa75ed71a1/language-en

[6] Majid,A. Pakistan's digital revolution is happening faster than you think. 2018. Available from: https://www.weforum.org/agenda/2018/11/pakistan-s-digital-revolution-is-happening-faster-than-you-think/

[7] Capobianco A, Nyeso A. Challenges for competition law enforcement and policy in the digital economy. Journal of European Competition Law & Practice, 2017; 9(1): 19-27.

[8] Caillaud B, Jullien B. Chicken & egg: Competition among intermediation service providers. The RAND Journal of Economics, 2003; 34(2): 309-312.

It does not encompass online markets developed by brick and mortar businesses to sell the products and services they produce. Digital platforms include activities and services such as search engines, social networking, payment systems, and marketplaces.

Digital platforms provide their services at zero costs in exchange user data [9]. They have developed complex algorithms to process the data, and decisions are based on that data. As such, the platforms require enormous initial investments or sunk costs but have low marginal costs [10]. This is because the technological systems required to collect, store, and process data can be costly, but once implemented, the marginal costs associated with additional data are low [11]. Due to the high sunk costs and thus, high economies of scale and scope, natural barriers to entry into digital markets exist and the market is concentrated within the hands of a few players, or superplatforms. Secondly, in multi-sided markets, services are provided for free to one customer group. The links created between two or more sides enables the platform to charge one group of customers and use the revenues to provide services to another user group for zero price [4]. This calls for legislative clarification whether such services fall under the purview of antitrust law.

Additionally, since there is an indirect network between suppliers of goods and services and consumers within the platform, then there exists an antitrust market regardless of whether one or more platform users are charged [10]. Another feature of online platforms is data-driven network effects [12]. This refers to the effect created by the use of one party on the value of the service to existing or potential users. For instance, a person may use Facebook or shop on Amazon because their friends do so. It, therefore, follows that the size and number of users determine the value of the platform. These cross-network effects create feedback loops that can make it difficult for new platforms with a small user base to compete against incumbents with millions or even billions of users [13].

---

[9] Hylton KN. Digital Platforms and Antitrust Law. *SSRN Electronic Journal*. 2019; 1-21.

[10] King SP. Sharing economy: What challenges for competition law? Journal of European Competition Law & Practice. 2015; 6(10): 729-734.

[11] United Nations Conference on Trade and Development (UNCTAD). *Competition Issues in the Digital Economy* (Report of the Intergovernmental Group of Experts on Competition Law and Policy, Eighteenth Session Geneva, 2019; 10–12 July

[10] King SP. Sharing economy: What challenges for competition law? Journal of European Competition Law & Practice. 2015; 6(10): 729-734

[12] Organization for Economic Cooperation and Development (OECD). Rethinking Antitrust Tools for Multi-Sided Platforms. 2018. Available from: https://www.oecd.org/competition/rethinking-antitrust-tools-for-multi-sided-platforms.htm

[13] World Economic Forum. *Competition Policy in a Globalized, Digitalized Economy*. 2019. Available from: https://www.weforum.org/whitepapers/competition-policy-in-a-globalized-digitalized-economy

The large economies of scope and scale, data control, and network effects create barriers to entry, a phenomenon which has facilitated the rise of giant digital companies such as Alibaba, Amazon, Facebook, and Google. New entrants lack the advantage of "big data" and this poses a great challenge for small companies in establishing large and successful platforms [14]. Furthermore, the dominant digital platforms prioritize user base maximization to profit maximization. Start-ups with insufficient capital are not able to sustain themselves to growth with such a business strategy. Even if they overcome these challenges, they are acquired by the incumbents before they can become potential competitive threats.

### 3. Challenges of Digital Platforms in the Context of Pakistani Competition Law

The first problem of tech companies to the antitrust framework in Pakistan relates to the determination of the relevant market. The purpose of market definition is to establish whether there is a need for ex-ante regulation, which may arise if a firm's activities are not constrained by potential or existing competitors, or if consumers cannot easily switch between competitive products. It provides a framework in which "the competitive aspects of anti-competitive agreements, abuses of dominance, mergers, or the need for regulation can be analyzed."

The Act defines the relevant market as "the market which shall be determined by the Commission with reference to a product market and geographic market. A product market comprises all those products or services that are regarded as interchangeable or substitutable by consumers because of the product's characteristics, prices, and intended uses. A geographic market comprises the area in which the conditions of competition are sufficiently homogeneous and which can be distinguished from neighboring geographic areas because, in particular, the conditions of competition are appreciably different in those areas." While this definition is inherently applicable to traditional business models, it is unclear whether each side of multi-sided platforms should be treated as a separate market. The relevant market for digital companies does not coincide with the technologies or products offered by the firms. Moreover, the innovative and dynamic nature of digital markets make it difficult to assess competitive restraints.

The Pakistani law deals only with demand-side substitutionality and makes no reference to supply-side substitution [15]. The consideration of potential supply-side substitutionality (whether supply by other firms is technologically feasible) is crucial in investigating abuse of dominance cases and in reviewing mergers.

---

[14] Maher M, Reynolds P, Muysert P, Wandschneider F. Resetting competition policy frameworks for the digital ecosystem. 2016. Available from: https://www.gsma.com/publicpolicy/resources/resetting-competition-policy-frameworks-for-the-digital-ecosystem

[15] Lundqvist B, Gal, MS. *Competition Law for the Digital Economy*. Gloucestershire, England: Edward Elgar Publishing. 2019.

Regarding demand-side substitution, changing customer preferences and the absence of well-established consumer views on product interchangeability make it challenging for competition authorities to determine which products and services are considered viable alternatives or substitutes [14]. As opposed to traditional industries where products with similar characteristics often comprise a market, the digital ecosystem is composed of a range of product and/or services which consumers may regard as substitutes [12]. Some competition authorities have stipulated guidelines in defining what constitutes a market for digital markets such as the U.S. Federal Trade Commission (FTC) guidelines which state that "the relevant market need not have precise metes and bounds [11]." The Pakistani competition law and guidelines have no such provisions.

The reliance on the price mechanism as an indicator of consumer welfare may also not be applicable to digital markets. Due to the significant network effects and economies of scale, the firms offer their services to consumers "free of charge" [15]. The other side of the market (the supply side – advertisers) bears the full burden of generating profits. Competition in digital markets is based on factors other than price such as product features, quality, and functionality. The traditional substantive tests cannot, therefore, be applied. The CCP should redefine the market by introducing guidelines that recognize the free products or services offered by online platforms. It is also noteworthy that since the platforms offer their services in exchange for consumer data, large-scale data gathering by players is a major source of market power and may have potential anti-competitive implications. The Commission should consider the significance of the scale and scope of data for competitive performance or institute data valuation mechanisms in its assessment.

---

[11] United Nations Conference on Trade and Development (UNCTAD). *Competition Issues in the Digital Economy* (Report of the Intergovernmental Group of Experts on Competition Law and Policy, Eighteenth Session Geneva, 2019; 10–12 July

[12] Organization for Economic Cooperation and Development (OECD). Rethinking Antitrust Tools for Multi-Sided Platforms. 2018. Available from: https://www.oecd.org/competition/rethinking-antitrust-tools-for-multi-sided-platforms.htm

[14] Maher M, Reynolds P, Muysert P, Wandschneider F. Resetting competition policy frameworks for the digital ecosystem. 2016. Available from: https://www.gsma.com/publicpolicy/resources/resetting-competition-policy-frameworks-for-the-digital-ecosystem

[15] Lundqvist B, Gal, MS. *Competition Law for the Digital Economy*. Gloucestershire, England: Edward Elgar Publishing. 2019.

Once firms in digital markets attain a dominant position, they have the ability to stifle innovation and competition by smaller firms within the same market by leveraging their market power [9]. Dominant firms, which often exhibit monopolistic tendencies, lock in end- users at both sides of the platform thereby making themselves indispensable [16]. Additionally, attempts by smaller firms to attain relevant market positions are thwarted by the incumbents through pre-emptive mergers – aimed at preventing the targets from becoming a competitive threat.

### 4. Merger Review in Pakistan

Mergers entail the restructuring of undertakings to achieve growth and efficiency for the benefit of consumers [12]. This is important for competition law and policy because the ultimate goal of competition law is to maximize consumer welfare. Thus, merger review constitutes a different class of competition law that often result in market detriment. The merger review process in Pakistan is ex-ante, implying that mergers are reviewed based on probabilities of their market effects [3]. Thus, at the time of review, these effects have not manifested themselves this presents a challenge. A poor evaluation of a proposed merger may lead to the clearance of a merger with anti-competitive effects on the market such as exclusionary effects or abuse of dominance if the post-merger entity is a dominant player. Also, a merger with potential benefits to consumers may be prohibited. The challenge of ex-ante review is even more present in the review of digital mergers [14]. Therefore, it is important for the competition authority to have a comprehensive understanding of the markets involved, and have sufficient tools at its disposal to effectively analyze digital mergers.

3 Wilson J. Crossing the Crossroads: Making Competition Law Effective in Pakistan. *Loyola University Chicago International Law Review*, 2011; *8*(2). Available from: https://lawecommons.luc.edu/lucilr/vol8/iss2/2/

[9] Hylton KN. Digital Platforms and Antitrust Law. *SSRN Electronic Journal*. 2019; 1-21.

[12] Organization for Economic Cooperation and Development (OECD). Rethinking Antitrust Tools for Multi-Sided Platforms. 2018. Available from: https://www.oecd.org/competition/rethinking-antitrust-tools-for-multi-sided-platforms.htm

[14] Maher M, Reynolds P, Muysert P, Wandschneider F. Resetting competition policy frameworks for the digital ecosystem. 2016. Available from: https://www.gsma.com/publicpolicy/resources/resetting-competition-policy-frameworks-for-the-digital-ecosystem

[16] OECD, Merger Control in Dynamic Markets. 2020. Available from: http://www.oecd.org/daf/competition/merger-control-in-dynamic-markets.htm

The Pakistani competition law captures only proposed mergers that meet the notification thresholds, based on turnover or asset value. In the digital industry where dominant incumbents acquire small fast-growing companies in different, the asset or turnover thresholds may not always be met and this results in the failure to review mergers that may have significant anti-competitive effects in the market on procedural grounds. As it will be discussed later in this paper, it is crucial that the CCP adopts transaction-based thresholds or issues guidelines that require mergers that do not meet the notification thresholds but which raise significant competition concerns be notified to the Commission.

5. **Data and Empirical Analysis**

   **5.1 CCP's Merger Review Activity**

   In our empirical analysis, we consider the mergers and acquisitions processed by the Commission since its inception in 2010 to 2019. The Commission processed a total of 710 applications between this period. Table 1 shows the number of M&A reviewed by the Commission on an annual basis.

   **Table 1:** Merger and acquisition reviewed by commission

| Year  | Phase I Review | Phase II Review | Total |
|-------|----------------|-----------------|-------|
| 2011  | 81             | 3               | 84    |
| 2012  | 73             | 4               | 77    |
| 2013  | 66             | 6               | 72    |
| 2014  | 81             | 5               | 86    |
| 2015  | 78             | 3               | 81    |
| 2016  | 64             | 5               | 69    |
| 2017  | 85             | 2               | 87    |
| 2018  | 68             | 5               | 73    |
| 2019  | 74             | 7               | 81    |
| **Total** | **670**    | **40**          | **710** |

The graph represents the CCP's performance for the period 2010 – 2019.

---

[11] United Nations Conference on Trade and Development (UNCTAD). *Competition Issues in the Digital Economy* (Report of the Intergovernmental Group of Experts on Competition Law and Policy, Eighteenth Session Geneva, 2019; 10–12 July

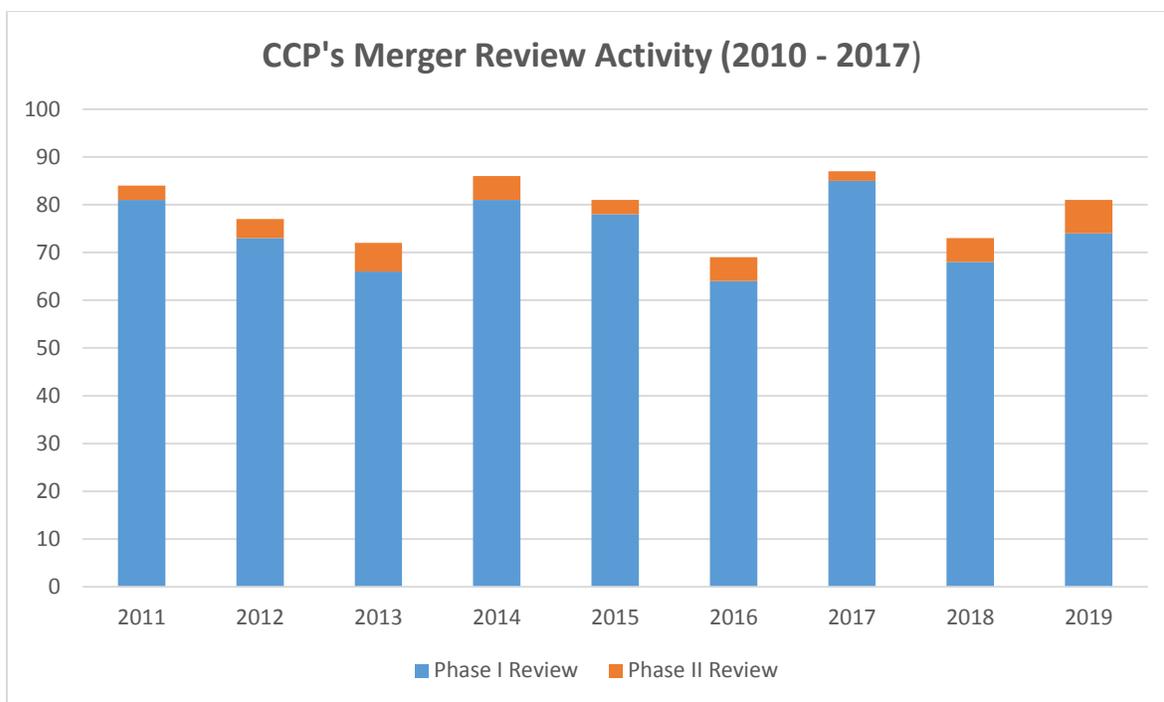

**Figure 1:** The graph represents the CCP's performance for the period 2010 – 2019.

The number of transactions approved without conditions, subject to conditions, and not approved are as shown in Table 2.

**Table 2:** Transactions approved with or without conditions

| Transactions permitted without conditions | 674 |
| --- | --- |
| Transactions subject to conditions | 36 |
| Transactions disapproved | 0 |
| **Total** | **710** |

## 5. 2 Classification Criteria of Digital Mergers

For the purposes of our analysis, a merger falls into the digital economy category if (i) it meets the specific criteria of digital platforms described above including but not limited to search engines, online marketplaces, ride-sharing applications, payment systems, and social networking; (ii) if both the acquirer and target are digital platforms. Based on these criteria, only 4 mergers in the digital economy were identified as provided in Table3.

**Table 3:** Identification of mergers in digital platforms

| Acquiring Firm | Target | Year of Acquisition | Approved without Conditions | Approved with Restrictions |
|---|---|---|---|---|
| Alibaba Singapore Holding (PVT.) LTD | JADE E-SERVICES PAKISTAN (PVT.) | 2018 | Yes | - |
| Hero | Food Panda Pakistan (PVT.) LTD | 2016 | Yes | - |
| Alibaba | Daraz Pakistan (PVT.) LTD | 2018 | - | - |
| Uber Inc. | Careem (PVT.) LTD | 2020 | Yes | - |

**5.3 Data Collection, Methodology, and Model Specification**

**5.3.1 Data Collection**

To compare the explanatory factors on M&A intervention by the CCP in the traditional and digital industries, we examined the public database of Pakistani mergers publicly available on the CCP's website (www.cc.gov.pk) covering 710 decisions from January 2011 to December 2019. Due to the small number of decisions in the digital sector, we constructed a random sample of 74 firms in the traditional industry to provide a more robust comparative model. All of the Commissions interventions in the digital economy were considered. The data includes the outcome of the CCP's final decision (dependent variable intervention). Since the outcome is ordered and discrete, we assign the arbitrary values of '0' if the CCP approved the merger without restrictions and '1' if the CCP disapproved the merger or approved with remedies regardless of whether the case involved a second phase review.

**5.3.2 Model Specification**

We estimate an ordered Probit model controlling for the year and month of decisions. A Probit model is expressed as $p = P[Z \leq \beta_0 + \beta_1.X] = F(\beta_0 + \beta_1.X)$ where xn are the explanatory variables and βn are the parameters of the explanatory variables, and with a probability curve p in the range [0,1]. Estimations are by the maximum likelihood method. The explanatory variables and parameters include industry, relevant market, geographic market, domestic acquirer ('1' if the acquirer is Pakistani and '0' otherwise), domestic target ('1' if target is Pakistani and '0' otherwise), barriers to entry, substitutes and existence of an undertaking. The CCP does not provide information about market shares and we have no way to obtain and quantify this information. Subsequently, the combined market share was omitted from the model.

The analysis uses two sets of data, distinguished by the acquiring industry. These include traditional economy sample (Model 1) and digital economy sample (Model 2). This classification enables the comparison of the factors that explain the Commission's decisions in the digital economy vis-a-vis the traditional economy. Our Probit model makes the following considerations:

(i)  If an explanatory factor is statistically significant for model 2 but is not for the other industries; and
(ii) If the factor is statistically significant for other industries but is not for the digital economy (model 1), there is evidence of distinction in the way the CCP deals with digital economy mergers.

### 5.4 Empirical Results and Discussion

Table 4 provides the descriptive statistics of the dependent and independent variables.

**Table 4:** Descriptive statistics of dependent and independent variables

|  | Relevant market | Geographic market | Barriers to entry | Substitutes | Existence of an undertaking | Domestic Target | Domestic Acquirer |
|---|---|---|---|---|---|---|---|
| Mean | 0.89 | 0.56 | 0.62 | 1.26 | 0.56 | 0.62 | 0.49 |
| Standard Error | 0.11 | 0.18 | 0.62 | 0.37 | 0.18 | 0.21 | 0.54 |
| Standard Deviation | 0.33 | 0.53 | 1.87 | 1.10 | 0.53 | 0.16 | 0.26 |
| Sample Variance | 0.11 | 0.28 | 3.49 | 1.21 | 0.28 | 0.32 | 2.58 |
| Minimum | 0.00 | 0.00 | -2.23 | -1.53 | 0.00 | -1.22 | 0.00 |
| Maximum | 1.00 | 1.00 | 2.45 | 2.22 | 1.00 | 1.00 | 1.45 |
| Confidence Level(95.0%) | 0.26 | 0.41 | 1.44 | 0.84 | 0.41 | 0.53 | 0.21 |

The results of our probit models are as shown in Table 5.

**Table 5:** Results of probit models

| Variable | Traditional Economy Sample | Digital Economy Sample |
|---|---|---|
|  | Model 1 (n=77) | Model 2 (n=3) |
| Relevant market | -0.02 | 0.01 |
|  | -0.05 | -0.11 |
| Geographic market | 0.2 | 0.04 |
|  | 0.03 | -0.02 |
| Barriers to entry | -0.05 | 0.28 |
|  | 0.44 | -0.05 |
| Substitutes | 0.00 | 0.02 |
|  | 0.00 | -0.01 |
| Existence of an undertaking | -1.39 | 0.16 |
|  | 0.42 | 0.15 |
| Pakistani Target | 0.07 | -0.16 |
|  | -0.25 | 0.37 |
| Pakistani Acquirer | -0.38 | -1.1 |
|  | -1.6 | -0.85 |

Model 1 reveals that relevant market ($\beta$=-0.02, p=-0.05), geographical market ($\beta$=0.2, p=0.03), barriers to entry ($\beta$=-0.05, p=0.44) and existence of an undertaking ($\beta$-1.39, p=0.42), Pakistani target ($\beta$=0.07, p=-0.25), and Pakistani acquirer ($\beta$=-0.38, p=-1.6) are statistically significant. Substitutes ($\beta$=0.00, p=0) are not statistically significant. For model 2, relevant market ($\beta$=0.01, p=-0.11), geographical market ($\beta$=0.04, p=-0.02), substitutes ($\beta$=0.02, p=-0.01) and existence of an undertaking ($\beta$=0.16, p=0.15), Pakistani target ($\beta$=-0.16, p=0.37), and Pakistani acquirer ($\beta$=-1.1, p=-0.85) are statistically significant.

The results show that the relevant market, geographical market, entry barriers and Pakistani target have significant explanatory power on the CCP's merger decisions regardless of the industry under analysis. The existence of an undertaking and Pakistani acquirer explain intervention in both the traditional and digital industries. While substitutes have little or no influential effect on merger decision outcomes in the traditional industry, they are relevant in determining the outcome of merger review decision in the digital economy. In sum, the results show no evidence on distinctive aspects that influence the CCP's decisions on merger reviews in the digital economy versus the traditional industries.

### 5.5 Correlation Matrix

To improve the reliability of our findings by avoiding multicollinearity bias, we compute the correlation matrix as shown in Table 6 below. It can be observed that the product market and geographic markets are highly correlated with the CCP's intervention in the intended merger. Similarly, barriers to entry, existence of an undertaking, and the nationality of the target (Pakistani or otherwise) also significantly influence the CCP's decision to intervene.

**Table 6:** Correlation matrix

|  | 1 | 2 | 3 | 4 | 5 | 6 | 7 | 8 |
|---|---|---|---|---|---|---|---|---|
| Intervention (1) | 1.00 | | | | | | | |
| Product market (2) | 0.29 | 1.00 | | | | | | |
| Geographic market (3) | 0.19 | 0.03 | 1.00 | | | | | |
| Barriers to entry (4) | 0.13 | 0.11 | 0.05 | 1.00 | | | | |
| Substitutes (5) | 0.05 | 0.01 | -0.04 | -0.03 | 1.00 | | | |
| Existence of an undertaking (6) | 0.12 | 0.08 | 0.40 | 0.27 | 0.16 | 1.00 | | |
| Pakistani Target (7) | 0.30 | -0.18 | 0.22 | 0.16 | -0.11 | -0.24 | 1.00 | |
| Pakistani Acquirer (8) | 0.00 | 0.02 | -0.19 | 0.05 | -0.06 | 0.04 | 1.00 | 1.00 |

### 6. Recommendations and Conclusion

The overarching goal of competition policies around the world is to prevent economic agents from abusing their economic power in order to maintain and enhance consumer welfare [11]. While most antitrust policies have been effective in curbing anti-competitive behavior for tangible goods and services markets, the new information economy has imposed challenges to many competition authorities including the Pakistani competition commission.

---

[11] United Nations Conference on Trade and Development (UNCTAD). *Competition Issues in the Digital Economy* (Report of the Intergovernmental Group of Experts on Competition Law and Policy, Eighteenth Session Geneva, 2019; 10–12 July

In recent months, a consensus has developed among competition law scholars, economists, and policy makers that the current anti-trust policy framework should be remodeled to effectively deal with the problems of platform ecosystems to strengthen competition in digital markets [9;15-16]. This paper investigates whether there is shift in how competition authorities in developing countries approach mergers in the digital sector by empirically evaluating whether the determinants of merger review decisions in the traditional economy also determine merger review decisions in the digital economy. More specifically: Are merger review decisions in the digital market by the CCP explained by the same explanatory factors as for the traditional industry? To answer this question, we examine seven potential concepts associated with the traditional theories of competition and regulation. We develop seven Probit regressions and a sample of 77 mergers in both sectors of the economy processed by the Commission between 2011 and 2017. Our analysis provides evidence that nearly all determinants (with the exception of substitutes) of merger decisions in traditional markets also inform merger review decisions in the digital platform market. In general terms, there is no distinction in how the CCP deals with digital mergers.

To ensure competitive markets across the economy, the Pakistani competition law and policy needs to adopt to the market realities of digital business models. The consumer welfare standard approach based on the price mechanism in insufficient, if not inapplicable, to digital platforms. UNCTAD [11] states that this framework ignores practices such as predatory pricing which are rampant in the digital sector. Dominant platforms, particularly such as Alibaba, Google, and Facebook, may lower their prices in the short to medium term, but once they achieve a monopoly position in the market, prices increase in an already limited choice environment. Additionally, personalized pricing by digital platforms and rapidly changing prices pose a difficulty in analyzing prices of online platforms providing marketplace infrastructure [15]. Furthermore, price analysis is not the most suitable standard of assessing the pro- and anti-competitive effects of digital platforms in the absence of given that they offer their services for free. Since, in fact, they provide services in exchange of consumers' personal data, digital mergers may result in consumer harm in forms other than price. As UNCTAD [11] posits, "consume welfare should be broadened to include other criteria such as consumer privacy and choice, personal data protection, switching costs, and the lock-in effects of online platforms."

[9] Hylton KN. Digital Platforms and Antitrust Law. *SSRN Electronic Journal*. 2019; 1-21.

[11] United Nations Conference on Trade and Development (UNCTAD). *Competition Issues in the Digital Economy* (Report of the Intergovernmental Group of Experts on Competition Law and Policy, Eighteenth Session Geneva, 2019; 10–12 July

[15] Lundqvist B, Gal, MS. *Competition Law for the Digital Economy*. Gloucestershire, England: Edward Elgar Publishing. 2019.

[16] OECD, Merger Control in Dynamic Markets. 2020. Available from: http://www.oecd.org/daf/competition/merger-control-in-dynamic-markets.htm

The increasing role of data as an important component of the digital market has also not been factored in Pakistani competition law. Digital platforms generate value from processing data collected "on the basis of free input of the platform users" [17]. They do not trade data as a stand-alone product but use to improve the services offered on the platform. For the digital industry, data is a form of "specialized asset," and competition policy should, therefore, regard it as sui generis information relevant for competition law purposes [5]. Pakistan has several laws in place that govern the collection, use, and protection of consumer data including the Personal Data Protection Act 2019 and the Prevention of Electronic Crimes Act 2016. There is also sector-specific legislation that impact data protection such as the Payment Systems and Electronic Funds Transfer Act 2007 that provides for the protection of customer information held by financial institutions and the Telecom Consumers Protection Regulations 2009, governing the use of subscribers' personal data. However, there are no legal provisions for the use of data by digital platforms. Moreover, enforcement of such provisions would require close collaboration between the CPP and the National Commission for Personal Data Protection. The CCP should develop guidelines or amend the **Merger Control Regulations (2016)** to include the protection and the control over the use of data by undertakings, ex post. Regulating data use in the industry with regard to the portability of data over platforms, open data, and revising the existing data protection legal frameworks currently in force would also promote the uptake of digital start-ups that rely on big data for growth.

The competition legislation also needs to redefine the relevant market for digital platforms to include factors beyond the product and geographic markets. With regard to the multi-sided nature of digital platforms, it is not clear whether the demand and supply sides should be treated as separate markets. Consumers, content providers, and advertisers do not engage with other on the platform due to different degrees of substitutionality on each side of the platform and this may give rise to consumer harm. For example, search engines and social networks may be viewed as substitutes by advertisers but are not by consumers. The current definition does not take into account the free nature of goods provided by digital platforms. Whole some European competition authorities such as Germany's Federal Cartel Office (FTO) have developed a digital-industry specific relevant market definition [18], developing countries such as Pakistan are yet to adopt such provisions. Concerning data, the antitrust law should also define a wider market for data. As earlier proposed, data is a form of specialized asset for digital platforms. Creating a "data market" would enable market players to compete for the asset that is used as an input to develop or improved the services provided by digital platforms.

---

5 Crémer J, Montjoye YA, Schweitzer H. Directorate-General for Competition (European Commission). Competition policy for the Digital Era. 2019. Available from: https://op.europa.eu/en/publication-detail/-/publication/21dc175c-7b76-11e9-9f05-01aa75ed71a1/language-en

17 Geradin D. What should EU competition policy do to address the concerns raised by the digital platforms' market power? SSRN Electronic Journal. 2018. 1-14.

18 Koenig C. Germany · Digital Economy, Antitrust Damages, and More: The 9th Amendment to the German competition act. European Competition and Regulatory Law Review, 2017; 1(3): 261-265.

As in many other jurisdictions, the Pakistani competition law requires only mergers fulfilling the turnover and/or asset thresholds to notify the commission of an intended merger. In the digital market, dominant platforms acquire small start-ups with low turnovers or small asset bases. Thus, while the start-ups are highly valuable, they may be by the CCP without the necessary scrutiny. The Commission should address these concerns. The FTO, for example, added a new threshold in the German competition law for notification requirements by proposed digital mergers. In Germany, in addition to the word wide turnover threshold and the first domestic turnover threshold, transactions whose consideration exceed 400 million euros should be notified to the FTO [18]. This captures the acquisition small innovative companies with less than the 5 million turnover domestic threshold but with high growth potential. In transactions where an incumbent acquires a quickly growing start-up that operates outside its market ("conglomeral transactions"), raises the question of whether stand-alone growth or M&A has more social value. Petit [19] proposes that the theory of harm should evolve to "think of the target firm as a horizontal competitor active within a same "technological space" or "user space." As such, if it can be proven that the target can grow as a competitive force by itself if not acquired by the incumbent, then an acquisition by the incumbent should be prohibited.

In conclusion, this paper provides empirical evidence that the Pakistani competition commission, in conducting merger reviews, does not treat digital mergers distinctively from mergers in the traditional industry. As theoretically evidenced in this paper, digital platforms are not only market disruptive but also raise important competition law issues that should be addressed. As UNCTAD [11] states, the ideal competition tools, including the competition authority, should detect and eliminate the potential competition restraints from mergers at the start rather than trying to correct anti-competitive outcomes ex post as the later may be difficult once a firm has already attained a monopoly position. Therefore, the CCP should push for legal amendments or develop separate competition policy guidelines for the digital market. As the Pakistani government spearheads a digital transformation campaign in the country, it is likely that tech start-ups will increase in the coming years. Changes to the competition law framework should go hand in hand with the digital transformation agenda.

---

[11] United Nations Conference on Trade and Development (UNCTAD). *Competition Issues in the Digital Economy* (Report of the Intergovernmental Group of Experts on Competition Law and Policy, Eighteenth Session Geneva, 2019; 10–12 July

[18] Koenig C. Germany · Digital Economy, Antitrust Damages, and More: The 9th Amendment to the German competition act. European Competition and Regulatory Law Review, 2017; 1(3): 261-265.

[19] Petit N. European Competition Policy in Digital: What's Next? - Competition Policy International. 2019. Available from: https://www.competitionpolicyinternational.com/european-competition-policy-in-digital-whats-next/

## 7. Limitations and Future Research Direction

Although this study is greatly important for competition law and policy for the digital sector in Pakistan, it has two major limitations which future research in the subject should address. First, the analysis did not include important factors associated with the competition policy theories. These include factors such as market shares and efficiency gains. This can be attributed to the use of secondary and public information. Thus, our models could not capture the relevance of these factors in explaining M&A interventions in the digital sector. In future, all the relevant factors for competition policy analysis should be considered. Secondly, the different sample sizes of the traditional and digital markets with a small size of the latter may have skewed the results towards the traditional market (Type II error) thereby decreasing the statistical power of the study and limit the generalizability of the sample to the digital sector population. A larger sample size for the digital and traditional sectors should be used in future to increase the generalizability of the results.

**Acknowledgements:**

.